\documentclass[aps, prd, twocolumn, lengthcheck, superscriptaddress, showpacs, nofootinbib]{revtex4-1}

\usepackage{epsfig}
\usepackage[usenames]{color}
\usepackage{graphicx}
\usepackage{amsmath}
\usepackage{epstopdf}

\def\bi{\bibitem}

\def\la{\langle}\def\ra{\rangle}
\def\be{\begin{eqnarray}}\def\ee{\end{eqnarray}}
\def\lsim{\mathrel{\rlap{\lower3pt\hbox{\hskip1pt$\sim$}}
     \raise1pt\hbox{$<$}}} 
\def\gsim{\mathrel{\rlap{\lower3pt\hbox{\hskip1pt$\sim$}}
     \raise1pt\hbox{$>$}}} 
\def\del{\partial}
\def\V{\cal V}\def\O{\cal O}

\allowdisplaybreaks

\begin{document}

\title{A Solution to the  Quenched  ${g_A}$ Problem \\ in Nuclei and Dense Baryonic Matter}

\author{Mannque Rho}
\email{mannque.rho@ipht.fr}
\affiliation{Institut de Physique Th\'eorique,
	CEA Saclay, 91191 Gif-sur-Yvette c\'edex, France }
\date{\today}

\begin{abstract}
When scale symmetry is combined with chiral symmetry in a scale-chiral Lagrangian, it can be shown in Fermi-liquid fixed point theory that  $g_A^{\rm eff}\approx 1$ in finite nuclei {\it as well as} in dense baryonic matter.  This is suggested as a signal for emergence of hidden symmetries of QCD in baryonic matter  from low to very high density. This calculation throws doubt on the ``first principles" explanation of the quenching of $g_A$ in nuclei with two-body meson-exchange currents. It also has relevance to Gamow-Teller matrix elements in neutrinoless double $\beta$ decay.
\end{abstract}

\maketitle

\section{\bf Introduction}
There is a long-standing ``mystery" lasting more than four decades~\cite{wilkinson} as to why the Gamow-Teller transition in shell model in nuclei seems to require a universal quenching factor $\sim 0.75$ multiplying the axial coupling constant $g_A$ measured in neutron decay, which would make the effective axial-vector  coupling constant $g_A^{\rm eff}\approx 1$. What was striking then -- and is more so now -- is that the resulting $g_A$ is surprisingly close to 1 in light and medium nuclei~\cite{quenching} updated in  the review~\cite{review-gA,review-doublebeta}. It could very well have been 2.0  or 0.5 or any other in that matter.  This prompted Denys Wilskinson from early 1970's~\cite{wilkinson}, and many others up to today since then,  to inquire whether this is not associated with something more than just mundane nuclear renormalization, something intrinsically tied to a basic property of QCD in nuclear medium. In the modern parlance, both the vector and axial vector currents are conserved if one assumes that the current quark masses for the up and down quark are zero. While the conserved vector current implies that the vector coupling constant $g_V=1$, the conserved axial current does not imply that $g_A=1$. In fact, there is nothing which says it should be even close to 1 even though  the current is exactly conserved. In Nature, it is $g_A\approx 1.27$. This is now understood as that axial symmetry is a hidden symmetry unlike the  vector symmetry which is unhidden.

So what does the axial coupling constant $g_A$ near 1 mean?

One way of seeing what it means is via the celebrated Adler-Weisberger sum rule which follows from the current algebras of chiral symmetry~\cite{adler,weisberger}
\be
g_A^2=1&+& f_\pi^2\frac{2}{\pi}\int^\infty_{m_N+m_\pi} \frac{WdW}{W^2-m_N^2}
\big[\sigma^{\pi^+ p} (W)-\sigma^{\pi^- p} (W)\big], \nonumber
\ee
with  the integral from threshold to infinity involving the difference of $\pi^{\pm} p$ scattering. This is on free proton, but one can imagine obtaining this sum rule on a nucleus $A$ taking the nucleus described as an elementary particle of spin 1/2.  This sum rule, if applicable to nuclei, then would give a simple answer: $g_A\to 1$ if either $f_\pi\to 0$ or the integral over the difference of $\pi^\pm A$ scattering vanishes. There seems to exist no obvious or solid reason why the difference should go to zero in finite nuclei while it does not on proton target to give $\sim 0.27$. The alternative with the pion decay going to zero is certainly a possibility since it is expected at some high density when chiral symmetry is restored. The difficulty here is that there is no reason why the pion decay constant should go to zero in $sd$-shell, $pf$-shell etc nuclei. There is an indication that it could be decreased at most by about $\sim 20\%$ in deeply bound pionic system.

This issue got highlighted recently by a remarkable ``work-of-the-art" computation of Gamow-Teller transitions in light and medium-nuclei, in particular in $^{100}$Sn~\cite{firstprinciple}. This work combines no-core shell model technique, thereby incorporating ``virtually exact" correlations in the nuclear wavefunctions, and EFT treatment of strong and weak interactions of the Standard Model.   Calculations along the similar line have been around since 1980, but what distinguishes this work form the previous ones  is the accuracy with which both high-oder nuclear correlations and effective field theory treatment of nuclear force and many-body weak currents are put together.  The calculation of \cite{firstprinciple} is focused on the super-allowed Gamow-Teller decay of the doubly magic $^{100}$Sn nucleus which exhibits the strongest Gamow-Teller strength so far measured in nuclei~\cite{Sn100},  an ideal system for large-scale calculation that can take into account a large number of particle-hole correlations.   The conclusion of this work is that in the state-of-the-art calculation in $^{100}$Sn combining  the ``virtually exact" correlations of many-body nuclear interactions anchored on chiral effective field theory on strong interactions and electroweak currents leads to the quenching factor $q=0.73-0.85$ which gives
\be
g_A^{\rm eff}= 0.95 - 1.08.\label{gAeff}
\ee

This calculation, as the title indicates, is heralded as a ``first-principles" resolution of the long-standing puzzle.

Now the question I would like to ask is to what extent this result  can -- or cannot -- be taken as a first-principles solution.  This issue is closely tied to whether the $g_A^{\rm eff}\simeq 1$ is (a) a coincidental outcome of  nuclear renormalizations or (b) a fundamental renormalization encoded in QCD or (3) a combination thereof. This is by itself an important issue for how QCD manifests in nuclear processes, but it is also a practically crucial element in addressing the neutrinoless double-$\beta$ process where such a quenching factor could play a significantl role~\cite{review-gA,review-doublebeta}.

In this note I address this question with an effective field theory that takes into account hidden symmetries of QCD, i.e.,  scale symmetry and hidden local symmetry, in a  chiral symmetry framework phrased in terms of Wilsonian renormalization-group approach to Landau(-Migdal) Fermi-liquid theory.

The answer I will arrive at is that $g_A^{\rm eff}\approx 1$ is given, in the vicinity of ordinary nuclear matter density ($n_0\simeq 0.16$ fm$^{-3}$), by combining two limits: QCD in the large $N_c$ limit in EFT and the Landau Fermi-liquid theory for baryonic matter in the large $\bar{N}$ limit where $\bar{N}=k_F/(\Lambda_{\rm FL}-k_F)$ ($\Lambda_{\rm FL}$ being the cutoff scale above the Fermi sea of the many-body system). I will explain why this is the {\it entire} result for the $^{100}$Sn GT beta decay.  It will also be argued that as density increases much beyond  that of the ordinary nuclear matter,  the ``fundamental" $g_A^{\rm fund}\to 1$ in the limit the dilaton decay constant $ f_\chi\propto \la\chi\ra$ encoding the scale symmetry breaking in dense medium goes to zero. The question then arises why and how $g_A^{\rm eff}\approx 1$ goes over to $g_A^{\rm fund}=1$.
\section{Nuclear EFT}
\subsection{Degrees of freedom}
The ``first principle" involved in the calculation of \cite{firstprinciple} is the standard chiral EFT that is to capture low-energy nonperturbative QCD  by including, as relevant degrees of freedom,  the pions in addition to the nucleons -- proton and neutron. I will  refer to this as S$\chi$EFT$_\pi$. Other degrees of freedom will be brought in to improve on S$\chi$EFT$_\pi$ in what follows later. For the moment, I will limit to this EFT since it is what is used in \cite{firstprinciple}.

That S$\chi$EFT$_\pi$ figures as a first-principles approach is along the line of reasoning given by S. Weinberg's Folk Theorem (on EFT) as applied to QCD. Apart from the symmetries etc. required, what is needed for nuclear EFT is the effective cutoff scale $\Lambda_{\rm eff}$ involved in nuclear dynamics. In the usual S$\chi$EFT$_\pi$ calculations practiced in the field as in \cite{firstprinciple}, the cutoff is taken $\Lambda_{\rm eff}\sim 400-500$ MeV. This means that  the vector mesons $V = (\rho, \omega)$ are integrated out of the meson sector as their free-space mass is greater than the cutoff $\Lambda_{\rm eff}$. Furthermore there is no scalar. This is for two reasons. First the  scalar $\sigma$ figuring  in relativistic mean field theories (referred in the literature as RMFT)\footnote{In order to avoid confusion, let me define the scalar involved in nuclear physics. Both the scalar in RMFT and the fourth component of the chiral four vector  in the linear sigma model are denoted in the literature as $\sigma$. They are not the same. The $\sigma$ that will figure later in scale symmetry is a dilaton, a Nambu-Goldstone mode of hidden scale symmetry, different from all others. It turns out however to be related to the one in the linear sigma model in some density regime, but they should not be confused.}  is of higher mass than $\Lambda_{\rm eff}$, so it is integrated out. Secondly it is considered to appear as a resonance appearing at high loop-orders in S$\chi$EFT$_\pi$, so should not be included in S$\chi$EFT$_\pi$.

As stressed by Weinberg, the nucleons with the mass $\sim 1$ GeV figure in nuclear EFT in the spirit of the Folk Theorem because what is involved in nuclear physics are ``soft" fluctuations comparable to soft pions in low-energy strong interactions.  Now the $\Delta (3,3)$ resonance is some $\sim 300$ MeV heavier, so one would think it could be left out of nuclear EFT,  S$\chi$EFT$_\pi$. This is what's done in \cite{firstprinciple}. However the mass difference $\sim 300$ MeV is comparable to the  energy of the nuclear states strongly excited  by the nuclear tensor force, and hence it seems unjustified to ignore the $\Delta$ degree of freedom. I  will come back to this matter later. For the moment I will continue ignoring the $\Delta$ considering that {\it it is integrated out}.
\subsection{Many-body currents}
In \cite{firstprinciple}, the nuclear forces ${\cal V}$ and the manny-body currents ${\cal O}$ are considered, respectively,  up to N$^4$LO for the former and up to N$^3$LO for the latter in chiral power counting. With the suitable $\V$ --  the reliability of which I will assume for the moment and to which I will return later -- the wavefunctions are {\it precisely calculated} given the powerful quantum many-body technique. This is considered as the fist step to what might be called a ``first-principles" calculation. There can be an objection here\footnote{An axiomatic theorist would raise a strong objection here. A strict consistency would require that given the EFT Lagrangian of QCD, the shell model should be ``derived" with the nuclear force given by the Lagrangian, say, as a sort of nontopological chiral soliton or something similar. This of course has not been done. For instance, getting Fermi surface in a system of interacting fermions  is a quantum critical phenomenon and using the EFT Lagrangian for fluctuations on top of given Fermi surface is already a hybrid approach with inevitable disregard of strict consistency. The same goes with the shell model. This is of course the lunatic axiomatic fringe but that is what a ``full consistency" with ``first-principles" would consist of. } but let us proceed assuming that this is OK although there is a caveat in connection with the cutoff scale involved in  $\V$ to which I will return  below.

 Now the issue of the many-body currents $\O$ needs  to be addressed.

 It turns out that unlike the vector currents that are quite straightforward the nuclear axial-vector currents turn out to be extremely  subtle.   This has to do with that the axial symmetry is ``hidden." The statement that it is ``spontaneously broken" is a misnomer.

 In fact it has been known since late 1970's that the space and time components of nuclear axial currents behave quite differently in nuclei and dense baryonic matter. This was evidenced in the current algebras before the advent of QCD in the way ``soft pions" come into two-body exchange currents~\cite{KDR}. In terms of the modern $\chi$EFT parlance, this is almost trivial. However the soft-pion theorems, just as all other soft theorems, be that photon or graviton, have a deep physical implication, ubiquitous in all areas of physics~\cite{soft}.
 \subsubsection{Protection by the chiral filter}
 What was clearly seen in \cite{KDR} in the absence of modern chiral counting in chiral Lagrangian was that the time component of two-body axial current is dominated by the exchange of a soft pion and could give $O(1)$ corrections to the first forbidden $A$-to-$B$ nuclear transition
 \be
 A(0^\pm)\to B(0^\mp) +e +\nu, \ \ \Delta T=1
 \ee
 where the superscripts are the parities.
 The prediction for nuclear matter~\cite{KR91} and the experimental measurements made for the transitions in Pb region $A=205 -212$~\cite{warburton} agreed stunningly well.
 \be
 \epsilon_{theory}=2.0\pm 0.2, \ \ \epsilon_{exp}=2.01\pm 0.05\label{a0}
 \ee
 where $\epsilon={g_A}^{\rm eff}_t/g_A$ expressed in terms of the effective axial coupling constant for the time component represents the ratio of the total matrix element over the single-particle matrix element.  The theoretical value is estimated at nuclear matter density, but the result is extremely insensitive to density, so Pb can be compared with nuclear matter: The 10\% error bar assigned to  the theory corresponds to the range of density involved from light to heavy nuclei to nuclear matter.  This shows that the two-body axial-charge matrix element contributes an equal amount as the leading-order (LO) single-particle one.
 \subsubsection{Non-protection by the chiral filter}
 The current algebra approach makes a clear prediction that the two-body soft-pion corrections to the space part of the axial current,  the Gamow-Teller operator, in stark contrast to the time component, are strongly suppressed. This dramatic difference was dubbed in 1970's as ``chiral-filter hypothesis," since at the time chiral perturbation theory was not yet around. Now there is a tool to justify the hypothesis. This hypothesis allows me to address the question raised above regarding the first-principles nature of the result of \cite{firstprinciple}.

 Let me first resort to the S$\chi$EFT$_\pi$ expansion for the axial current which was first derived early in 2000 and completely listed in 2003~\cite{parketal}. It has since been extensively refined and extended since then as summarized -- with relevant references -- in  \cite{krebs}. What I present below is essentially all contained in \cite{parketal}.

First look at the time component of the axial current. Here soft pions predominantly enter in the two-body current. The ratio of two-body soft-pion exchange operator over the one-body operator -- which is $O(Q)$ in the power counting -- is $R=$2B/1B$=O(Q^0)$. Thus the leading ``correction" is of the same magnitude as the LO one-body term. The next correction is strongly suppressed, say, by  two chiral orders -- $O(Q^2)$ in the power counting. At this order there are  relativistic and other corrections  to the single-particle operator as well as two-body terms involving 2$\pi$ exchange etc. Thus the leading two-body term is protected by  the ``chiral filter" and robust. One could say that the dominance of the soft pions in (\ref{a0}) makes  the Folk Theorem cleanly ``proven" in nuclear physics. As far as I know, this is the most convincing  -- and clear-cut -- evidence for the role of pions -- via exchange currents --  in nuclear physics. It would, of course, be highly valuable to confirm by high-power quantum many-body techniques that the prediction in (\ref{a0}) is not modified by higher-order corrections arguably strongly suppressed.

In stark contrast, the situation with the Gamow-Teller operator, the space component of the axial current, is drastically different, aptly dubbed the ``other side of the same coin." This is because soft pions play practically no role here.  While the one-body Gamow-Teller operator is $O(Q^0)$ and super-allowed -- barring  accidental suppression --  the leading two-body correction with one-pion exchange comes suppressed by two chiral orders,  $O(Q^2)$, so the ratio is $R=$2B/1B$=O(Q^2)$. This is because the pion entering in the two-body term is not soft, with its coupling with nucleons requiring, among others, relativistic corrections. At this order, three-body operators also enter. Furthermore since the nucleons are inevitably non-relativistic, there can be a multitude of  other corrections including  ``recoil corrections" entering at the same  order. It is not clear whether all these corrections are fully taken into account in \cite{firstprinciple}. There is no justification to take  some but ignore others as there can be significant cancellations among them. They may also be all essential for axial Ward identities.  At this chiral order, as pointed out in \cite{krebs}, there are also ambiguities in doing regularizations\footnote{This ambiguity is highly relevant to the validity or meaningfulness in correlating the presence of 2BC with the regularization (cutoff dependence,   a.k.a. ``resolution scale" etc.) discussed in \cite{firstprinciple}.} in both $\V$ and $\O$.  This plethora of uncontrollable higher-order terms is what is meant by ``chiral-filter unprotected" terms.

\begin{figure}[h]
\begin{center}
\includegraphics[width=6cm]{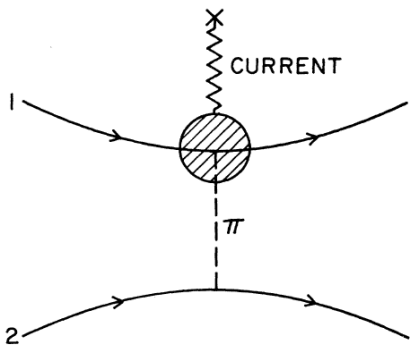}
\end{center}
\caption{Two-body exchange current. The upper vertex involves two soft pions for the axial charge transition.}\label{pi}
\end{figure}

In fact one can make a simple, though heuristic, argument to suggest that the one-pion exchange current to the Gamow-Teller transition cannot be important, if not completely ignorable,  for the problem. Consider the  vertex ${\cal A}_\mu +N\to \pi+N$ in the one-pion exchange graph in Fig. \ref{pi} where ${\cal A}_\mu$ is the external axial field.
Consider the axial field ${\cal A}_\mu$ as a pion. Then one is considering the process $\pi_{\rm in}+ N\to \pi_{\rm out} +N$. (a) Suppose $\pi_{\rm in}$  is ``hard" and $\pi_{\rm out}$  is ``soft."  Then according to the double soft theorems~\cite{soft},  the amplitude should be highly {\it suppressed} by Adler's theorem. (b) On the other hand if both $\pi_{\rm in}$  and $\pi_{\rm out}$  are soft, then the double-soft limit gives a  unsuppressed ($\sim O(1)$) amplitude. This is very well known from the old soft-pion theorems, but nowadays this old stuff has become highlighted  because of its fundamental nature in physics~\cite{soft}.  Kinematics in pion fields in nuclei is not  so well defined, so the argument is at best approximate. But with the axial current identified with a pion, this soft-theorem can be applied to the problem. The pion exchanged between two nucleon is favorable for the process when it is soft, with harder pions suffering from kinematic suppression due the derivative coupling. Now one can take the axial-charge current for small momentum transfer to be soft, whereas the axial-vector part  is hard. Thus the pionic 2BC for the Gamow-Teller transition should be {\it suppressed} whereas the 2B axial charge operator could be {\it enhanced}.  This was the content of the old chiral filter argument~\cite{KDR}.  The result (\ref{a0}) confirms (b). Now I am going to argue that (a) will also be confirmed.

As a summary of the discussion given so far, I would argue that the conclusion of \cite{firstprinciple} -- that the $g_A$ problem is resolved by the  2BC combined with a sophisticated no-core shell model -- is too hasty or even questionable. Apart from the caveat pointed out by \cite{krebs} -- which can be serious particularly with the chirally suppressed terms -- compounded with the adjustment of the cutoff scale, a.k.a. the  ``resolution scale,"  to maximize the 2BC, there is no justification to stop at N$^3$LO unless N$^4$LO can be shown to be ignorable, which is unfeasible at present. There can very well be cancelations between different orders as in the case of the Monte Carlo calculations in light nuclei~\cite{wiringa}.

I now present an approach to the problem that supports the assertion that {\it the N$^3$LO corrections (11 terms in \cite{krebs} plus other terms such as  recoil terms which may or many not have been included in \cite{firstprinciple}) cannot be the source for the resolution of the $g_A$ puzzle in general and in $^{100}$Sn in particular.}

\section{Scale-invariant hidden local symmetric EFT}
Here I will describe how one can calculate $g_A^{\rm eff}$ in effective field theory for strongly correlated baryonic matter following a Wilsonian RG approach~\cite{shankar}. I will do this in the limit $\bar{N}\to \infty$  referred to as ``Fermi-liquid fixed point (FLFP)" limit. The effective Lagrangian is defined with a cutoff put just above the vector-meson mass $\sim 700$ MeV, with the vector mesons $\rho$ and $\omega$ brought in as hidden gauge fields and a scalar corresponding to $f_0(500)$, denoted $\chi$ (to be distinguished from $\sigma$ of linear sigma model and also from the scalar in RMFT)  as a (pseudo-)Nambu-Goldstone scalar boson of scale symmetry. Those, in addition to the nucleons, are the relevant degrees of freedom for the given cutoff. There is no need for ``resolution-scale" adjustment. I will leave out the $\Delta$ for the moment and later argue that it is justified to ignore it. This EFT Lagrangian will be referred to as ``$bs$HLS."\footnote{Here $b$ stands for baryon, $s$ for scalar dilaton and HLS for hidden local symmetry fields.} Since in the large $N_c$ limit, $g_A$ goes  $\sim O(N_c)$, I will limit to $O(N_c)$ in computing $g_A^{\rm eff}$. Therefore the calculation I will do is valid for the limit of {\it large $N_c$ and large $\bar{N}$}.

\subsection{ RMF theory with $bs$HLS Lagrangian and Fermi-liquid fixed point}
The principal tool for the calculation  is the $bs$HLS Lagrangian with its bare parameters endowed with non-perturbative inputs in terms of condensates inherited from QCD at the matching scale between the EFT and QCD.  The explicit form of this Lagrangian  is given in detail in \cite{MR-review}, which was written for the primary purpose to describe how the EFT Lagrangian can be applied to dense compact-star matter. It turns out  that the review does contain what's needed for the present problem. The logic is rather involved and the details can be skipped for the discussion to be given here. I will just give the essential structure relevant to the problem at hand and present drastically simplified but what I consider to be correct arguments to give the key ideas and results.

As explained in \cite{MR-review},  HLS encodes chiral symmetry in terms of the vector mesons through gauge equivalence  to non-linear sigma model -- the basis for $S\chi$EFT$_\pi$ --  with the hidden gauge symmetry playing an extremely important role both at nuclear matter density and at high compact-star density. The dilaton  $\chi$ encodes the scale symmetry of QCD which is hidden in the vacuum. The scale symmetry and HLS, treated on the same footing, give rise to scale-chiral symmetry which is taken as the basis of nuclear strong dynamics in lieu of  chiral symmetry alone. The presence of the dilaton as an active degree of freedom makes the theory a lot more powerful -- and simpler --  than chiral symmetric theory, but it is extremely subtle and up-to-date barely developed.  It is clear even at a superficial level that it makes a great sense because what arises at very high loop orders in S$\chi$EFT$_\pi$ can be captured economically at tree order\footnote{One can see this already in particle physics, for instance in $K\to 2\pi$ decay~\cite{CT}.}.

Without getting into detailed expression, the $bs$HLS Lagrangian that will be employed is -- schematically -- written as
\be
{\cal L}_{bsHLS}={\cal L}_{inv} (\psi, U, \chi, V_\mu) + {\cal V} (U,\chi, {\cal M})\label{LOSS}
\ee
where the first term is scale-invariant and the second is the dilaton potential that encodes scale-chiral symmetry breaking.  Here $\psi$ is the nucleon field,  $U=e^{2i\pi/f_\pi}$ is the chiral field, $\chi=f_\chi e^{\sigma/f_\chi}$ is the conformal compensator field for the dilaton,  $V_\mu$ is the hidden gauge field. The dilaton potential puts the system in Nambu-Goldstone mode of scale-chiral symmetry.  The HLS is assured with hidden gauge covariance put in the Maurer-Cartan 1-forms and can be written down to any power orders.

I should mention  that how scale symmetry figures in pre- and post-QCD has a long history and is still highly controversial. It is currently also a hot topic in connection with the structure of Higgs boson and for the attempt to go beyond the Standard Model. As explained in \cite{MR-review,CNDIII}, the strategy used in accessing compact-star physics is based on the notion put forward by Crewther and Tunstall~\cite{CT} that for QCD with three flavors (u, d, s), there is an infrared fixed point for the QCD $\beta$ function with an $\alpha_s$ ``freezing" at $\alpha_{IR}$ far away from asymptotically free coupling, $\beta (\alpha_s=\alpha_{IR})=0$. The $f_0(500)$ is identified as the dilaton living in the vicinity of the IR fixed point.
It has been suggested that the Crewther-Tunstall (CT)  scheme is the most appropriate one for treating scale symmetry in dense medium, together with the hidden gauge symmetry emergent at high density. This matter is treated in detail in \cite{MR-review}. Whenever the occasion arises, I will point out in what way the CT scheme differs from other schemes proposed in the literature, in particular those schemes addressing dilatonic Higgs. The $bs$HLS Lagrangian (\ref{LOSS})  corresponds to what was referred in \cite{MR-review} to as ``leading-order scale symmetry approximation" to the CT theory.

Now the question to raise is: What does the matching with QCD at the matching scale do to the Lagrangian (\ref{LOSS})?

The matching is usually done at the chiral scale $\Lambda_{\rm chiral}\sim 1$ GeV. It is a bit above the cutoff picked for $bs$HLS, but that's the relevant scale for the degrees of freedom taken into account in $bs$HLS. . At that scale, the parameters of the EFT Lagrangian will have the dependence on various nonperurbative quantities, principally, the condensates, $\la\bar{q}q\ra$, $\la G_{\mu\nu}^2\ra$ etc., inherited from QCD. Embedded in the medium with a density $n$, then the dilaton potential picks up the dilaton condensate $\la\chi\ra$ so the $\chi$ field shifts
\be
\chi\to \la\chi\ra_n +\chi^\prime.
\ee
This then gives the bare parameters of the Lagrangian $n$ dependence. This dependence is called ``intrinsic density dependence (IDD)" in \cite{MR-review,CNDIII}. Now it is at this point that the in-medium dilaton decay constant $f_\chi^\ast$ gets locked to the pion  decay constant $f_\pi^\ast$, $f_\chi^\ast\approx f_\pi^\ast$. It should be made  clear that the IDD is in principle distinguishable from the density dependence that arises from mundane nuclear many-body interactions.

One can see from the explicit expression of the Lagrangian (\ref{LOSS}) that to the lowest order in scale-chiral expansion,   the Lagrangian has the form of Walecka's linear mean-field model~\cite{walecka}. The major difference from Walecka's model  however is that  the bare parameters,  endowed with the IDDs, are constrained by hidden local symmetry (hence chiral symmetry) and scale symmetry (hence low-energy theorems with the dilaton)  and of course the pion fields included \`a la nonlinear chiral symmetry. The RMFT that belongs to the class of energy density functional theory, often used in the nuclear physics community with large success, simulates to certain extent the effect  of IDDs by multidimensional field operators,  improving on certain properties of the linear model that are defective, e.g., the compression modulus of nuclear matter. It however does not possess the hidden symmetry properties of the EFT with $bs$HLS.

Now the most crucial observation is that this $bs$HLS Lagrangian treated in the mean-field can be taken to be equivalent to the Landau Fermi-liquid fixed point theory. The ``equivalence" was suggested a long time ago by Matsui for Walecka's linear RMFT~\cite{matsui}. Although a rigorous proof is sill lacking, there is a good indication that the RMF approach with $bs$HLS Lagrangian fairly closely represents  Landau Fermi-liquid point approximation~\cite{MR-review}. That the mean-field result of (\ref{LOSS}) is fully consistent with Fermi-liquid theory was shown in detail by Song in his thesis~\cite{song}. There it was shown that the proper treatment of the IDDs as a part of EFT is indispensable for thermodynamic consistency of the RMT.

One can clearly see this ``equivalence" in the nuclear response function to the EM field in comparing with Migdal's finite Fermi-liquid theory~\cite{migdal}. For instance  the Migdal formula for the orbital current
\be
\vec{J}=\frac{\vec{k}}{m_N} g_l
\ee
where
\be
g_l &=& \frac{1+\tau_3}{2} +\delta g_l\\
\delta g_l &=& \frac 16 (\tilde{F}_1^\prime -\tilde{F}_1)\tau_3\label{deltagl}
\ee
where $\tilde{F}_1$ and $\tilde{F}_1^\prime$ are Landau-Migdal interaction parameters
 is exactly reproduced by the MFT of (\ref{LOSS}) coupled to the EW fields. It satisfies the famous Kohn theorem, namely,  the appearance of the vacuum value of the nucleon mass $m_N$ instead of the the Landau mass, $m_L$, as  required by the current conservation and predicts the anomalous orbital gyromagnetic ratio (\ref{deltagl}) for proton with the IDD parameters of the Lagrangian (\ref{LOSS})~\cite{song}
\be
\delta g_l^p (n_0)\simeq 0.21
\ee
which is confirmed by what's measured in the Pb region, $\delta g_l^{\rm proton} = 0.23\pm 0.03$~\cite{schumacher}.\footnote{ As far as I am aware, this quantity fails to be explained by  S$\chi$EFT$_\pi$. The calculation that I am familiar with~\cite{HKW} gives $0.07\pm 0.02$, far short of the experimental value. It would be interesting to understand why the approach S$\chi$PT$_\pi$ (where the vector mesons and the dilaton are integrated out) fails so badly. } This is a ``back-of-envelope" proof that the MFT makes a good sense at least at the level of low-energy theorems.

Now having been assured that the vector low-energy theorems are encoded in this $bs$HLS theory, let me turn to the $g_A$ problem, which is linked to low-energy theorems in the axial channel that are  somewhat more intricate.

What I would like to do is to calculate the quenching factor $q$ associated with the $B_{\rm GT,ESPM}$ in $^{100}$Sn~\cite{Sn100}.

The quenching factor in the Fermi-liquid fixed point theory is  $q_{L}=g_{A}^{L}/g_A$ where $g_A^{L}$ is what corresponds to the Fermi-liquid fixed point constant that multiples the zero-momentum-transfer matrix element ${\cal M}=(\sum_i \tau_i\sigma_i)_{QP}$ for the quasi-particle on top of the Fermi surface making the GT transition, $M_{GT}=q_L g_A {\cal M}$. This $q^L$ should be compared with the experimental value
$q$ in $^{100}$Sn.

The relevant part of the Lagrangian (\ref{LOSS}) for this problem is
\be
 {\cal L} &=&i\bar{\psi} \gamma^\mu \del_\mu \psi -\frac{\chi}{{f_\chi}}m_N \bar{\psi}\psi +g_A \bar{\psi}\gamma^\mu\gamma_5 \tau_a \psi{\cal A}_{\mu}^a+\cdots
 \label{LAG}
 \ee
 where ${\cal A}_\mu$ is the external axial field.  The part given by $\cdots$ is not directly relevant. The key point  to note here  is that the axial response term in the Lagrangian is scale-invariant without dependence on the conformal compensator field $\chi$, whereas the nucleon mass term is linear in $\chi$. This means that embedded in nuclear medium, $g_A$ as a bare parameter is free of  IDDs, whereas the nucleon mass does scale ``intrinsically." Thus embedded in medium the Lagrangian takes the form
 \be
 {\cal L}^\ast &=&i\bar{\psi} \gamma^\mu \del_\mu \psi -m^\ast_N \bar{\psi}\psi +g_A \bar{\psi}\gamma^\mu\gamma_5 \tau_a \psi{\cal A}_{\mu}^a+\cdots
 \label{LAG-star}
 \ee
 where
 \be
 m_N^\ast &=&\Phi m_N, \\
 \Phi &\equiv & f_\chi^\ast/f_\chi\approx f_\pi^\ast/f_\pi\label{phi} .
 \ee
 The second approximate equality in (\ref{phi}) follows from the locking of the pion decay constant to the dilaton decay constant.  Note that this is a consequence of the same $N_c$ dependence of $f_\pi^2$ and $f_\chi^2$ in the CT theory.\footnote{I should mention that  this feature is not shared by other scale-symmetry theories. For example, there are certain theories in which the two decay constants have different $N_c$ dependence~\cite{GS} or have vastly different numerical values, $f_\pi/f_\chi \ll 1$~\cite{appelquist}. It turns out that the locked constants seem more or less consistent with nuclear dynamics, indicating that the CT scheme is favored in nuclear physics.}

 It should be strongly emphasized  that  while $g_A$ has no intrinsic density dependence, $f_\pi^\ast$ is directly affected by the IDD because of the locking to $f_\chi^\ast$, Eq. (\ref{phi}). This was already noticed in the Adler-Weisberger sum rule.

 At the Fermi-liquid fixed point, the relevant quantities involved are the Landau mass $m_L$,  the Landau interaction parameters  $\tilde{F}_1$ and $\tilde{F}_1^\prime$ and $\Phi=f_\pi^\ast/f_\pi$. Thus the Landau $g^L_A$ -- hence $q^L$ -- must involve only these quantities. The calculation for $g_A^L$ was first done a long time ago~\cite{FR}\footnote{In this reference, an argument was made using the Skyrme model with the scale invariant quartic term with the coefficient $1/e^2\sim O(N_c)$  which leads to  $g_A\sim O(N_c)$. The result is the same as is obtained now with (\ref{LOSS}) in the mean field.}. It is given by
\be
g_A^{\rm L}/ g_A \approx (1-\frac 13 \Phi\tilde{F}_1^\pi)^{-2}\label{gAL}
\ee
where $\tilde{F}_1^\pi$ is the pion Fock term contribution to the Landau parameter $\tilde{F}_1$. The Fock term is a loop contribution, so naively $O(1/\bar{N})$. But the pion being ``soft," it plays an indispensable role as it does for the anomalous orbital gyromagnetic ratio $\delta g_l^p$.  Since pionic properties are given by chiral dynamics, the pion contribution  $\tilde{F}_1$ can be calculated almost exactly. Thus  once $\Phi$ is given, then $g_A^L$ is accurately calculable.   How the pion decay constant behaves in nuclear medium is experimentally measured~\cite{yamazaki}, so $\Phi$ is known in the vicinity of nuclear matter density. Quite surprisingly while $\Phi$ decreases as density increases, the pionic term $\tilde{F}_1^\pi$ increases  with  the product $\Phi\tilde{F}_1^\pi$ staying nearly constant as density changes, say,   between $\sim \frac 12 n_0$ and $\gsim n_0$. Therefore $g_A^L$ is nearly density-independent, which predicts that the quenching factor $q^L$ must be more or less the same from light nuclei to heavy nuclei (and dense matter $n\gsim n_0$). The result is (evaluated at $n_0$)
\be
q^L\simeq 0.79.\label{qL}
\ee
This is essentially what's given in $^{100}$Sn~\cite{Sn100},
\be
q_{exp}=0.75(2).
\ee
As already alluded, what's surprising is that the Fermi-liquid fixed point result (\ref{qL}) which simply ignores $1/\bar{N}$ corrections, gives
\be
g_A^{\rm eff}\simeq 1.0.
\ee

What is significant of the result (\ref{gAL}) is that whereas $g_A$ at the matching scale, being scale-invariant,  has no dependence on IDD, the effective $g_A$ in nuclear matter $g_A^{\rm eff}$ is {\it crucially} dependent on it.  This comes about because the Landau mass of the nucleon does depend on it and the IDD dependence sneaks into $g_A^{\rm eff}$ through strong nuclear correlations involving the Landau mass. {\it It cannot be through the suppressed 2BC.}

\subsection{$g_A=1$ and the dilaton limit}
Let me describe one more surprising thing that is predicted by the $bs$HLS Lagrangian (\ref{LOSS}).

Instead of going to the Fermi-liquid fixed-point limit, let me take what is called ``dilaton-limit fixed-point" first considered by Beane and van Kolck~\cite{beane-vankolck},  which corresponds to taking the dilaton decay constant $f_\chi\sim \la\chi\ra$ going to zero~\cite{MR-review}. At that limit scale symmetry will be restored, that is, the system is driven to the IR fixed point $\beta(\alpha_{\rm IR})=0$. Exactly how that limit can be reached is far from clear. Some possibility is discussed in \cite{MR-review} in compact stars at high density.   Here I discuss how the matter close to it -- but not on top of it--  looks like.

To see what happens in that limit, take the Lagrangian (\ref{LOSS}) and do the field re-parametrization,
\be
\Sigma=U\chi\frac{f_\pi}{f_\chi}.
\ee
Now if one dials ${\rm Tr} (\Sigma\Sigma^\dagger)\to 0$ in the re-parameterized Lagrangian, there results a singular part
\begin{eqnarray}
\mathcal{L}_{\rm sing} & =&
\left( 1 -g_A \right) {\cal A} \left( 1/\mbox{Tr} \left[ \Sigma \Sigma^{\dagger} \right]\right) \nonumber\\
&&{} + \left( \delta -1\right) {\cal B} \left( 1/\mbox{Tr} \left[ \Sigma \Sigma^{\dagger} \right]\right)\,,
\label{sing}
\end{eqnarray}
where $\delta \equiv \frac{f_\pi^2}{f_\chi^2}$. Here ${\mathcal A}$ and ${\mathcal B}$ are singular quantities coming, respectively, from the baryonic  and mesonic Lagrangians. The requirement that ${\mathcal L}_{\rm sing}$ be absent leads to the conditions that
\be
g_A\rightarrow 1,\label{ga}
\ee
and
\be
f_\pi \to f_\chi .\label{locking}
\ee
Taken in the mean field, the quantities involved will be in-medium quantities and hence should be affixed with $\ast$.
Thus as density goes high, Eq. (\ref{ga}) predicts that the $g^\ast_A\to 1$ and  Eq. (\ref{locking})  predicts $f_\pi^\ast\leftrightarrow  f_\chi^\ast$. The latter again indicates the locking of the two decay constants encoded in the CT scheme~\cite{CT}.
It is intriguing that $g_A^{\rm eff}\to 1$ both at low density and at high density. What does this mean?
\section{Concluding remarks}
My conclusion is this: The Gamow-Teller matrix element in the beautiful  $^{100}$Sn beta decay is, most likely,  entirely governed by nuclear correlations with no intrinsic QCD scaling, in contrast to the pion decay constant, and no multi-body currents -- and if any, possibly tiny,  corrections from the $\Delta$s. This is in stark contrast to the axial-charge matrix element which is entirely governed by the well-controlled one-body and soft-pion two-body currents with little, if any, nuclear correlations. What underlies this result is the powerful double soft theorem for the pion.

How I arrive at this conclusion is as fallows.

In an effective field theory addressed to finite nuclei as well as to dense baryonic matter that implements a possible hadron-quark continuity in terms of hidden symmetries of QCD, the long-standing ``quenching factor" for $g_A$ in nuclear Gamow-Teller transitions $q\approx 0.73$, nearly independent of density,  can be explained {\it within the nucleon space} with one-body Gamow-Teller operator {\it alone} in the framework where hidden scale symmetry with the dilaton and hidden local symmetry with the vector mesons are combined.  In the large $N_c$ limit and Landau(-Migdal) Fermi-liquid fixed-point limit, the quenching is shown to arise entirely from strong nuclear correlations crucially tied to the nuclear tensor force as I will explain below. This implies that the many-body meson currents play no significant role due to the ``chiral filter" mechanism. The quantum Monte Carlo calculation in $A=6$-10 nuclei~\cite{wiringa} gives the results that are consistent with this conclusion.

It does not of course exclude the possibility that one could arrive at a correct quenching by what is called ``first-principles" calculation of the type discussed  in \cite{firstprinciple}. However given that the leading correction to the single-particle Gamow-Teller operator in the standard chiral power counting is suppressed strongly -- unprotected by the chiral filter --  there is nothing to suggest that the large number of un-calculable next-order terms can be ignored. Perhaps one should develop a basically different chiral ordering strategy.

A matter that was left out of my calculation is  the possible role of the $\Delta$ resonance in the quenching, which in the past was considered seriously in view of its important role of the resonance in pion-nuclear interactions~\cite{pion-nuclear}. I will now argue that the $\Delta$s cannot seriously affect  my conclusion either.

To bring out the point, first recall that Gamow-Teller states with the  excitation energy of $\sim 200-300$ MeV are strongly excited by the nuclear tensor force. The $\Delta$-hole states of comparable excitation energies due to the $\Delta$-N mass difference can also be excited in the Gamow-Teller channel by the tensor force acting between nucleons and $\Delta$s. This means that a powerful quantum many-body technique purported to give an explanation of the quenching factor $q$ must include both particle-hole and $\Delta$-hole states up to the excitation energy $\sim 300$ MeV.  This aspect is missing in the calculations so far done in the field.

To do this sort of calculation reliably, one would have to take into consideration the structure of the tensor force that predominantly excites the Gamow-Teller states. Apart from the different couplings involved with the nucleon and the $\Delta$, the structure of the tensor force is essentially the same for both the NN and N$\Delta$ channels, so let me discuss the NN channel. In fact this matter turns out to be highly important particularly in compact-star physics where high density is involved~ \cite{MR-review}.  Let me summarize what is relevant to the problem we have here.

\begin{figure}[ht!]
\begin{center}
\includegraphics[width=7.cm]{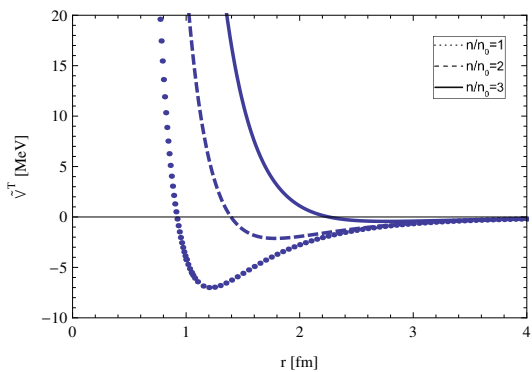}
\caption{Net tensor force $\tilde{V}^T\equiv (\tau_1 \cdot \tau_2 S_{12})^{-1} (V_\pi^T +V_\rho^T)$ in units of MeV.}\label{tensor}
\end{center}
\end{figure}

In the EFT anchored on $bs$HLS, the tensor force is given entirely by the sum of one-pion exchange and one-$\rho$ exchange forces. With the IDDs suitably incorporated, the net tensor force tends to drop in the strength in attraction as the density of the matter goes up. The reason for this drop is due to the cancellation between the two tensor forces.  Fig. \ref{tensor} illustrates how the tensor force loses attraction as density goes up from the nuclear matter density to 3 times the nuclear matter density. At $\sim 3n_0$, the tensor attraction nearly disappears, a feature that has the most drastic effect on the EoS for compact stars~\cite{MR-review}.

In the quenching problem, one is dealing with density only up to $\sim n_0$. However this continuous drop in attraction in the tensor force has a very important impact on certain nuclear processes, at a density $n\lsim n_0$, a spectacular case being the long C14 half-life~\cite{holt}\footnote{The reference \cite{firstprinciple} cites [18] and [19] for this process, ignoring \cite{holt} which I think is a lot more interesting. In both cases, the underlying mechanism is the same, namely the short-range three-body force.  In \cite{holt}, the three-body force involving $\omega$-exchange(s) is integrated out,  with part of its effect going into the  coefficient of the two-body tensor force. It is the scalar dilaton condensate that is responsible for this effect. This illustrates how the dilaton scalar brings a higher-order term into a lower-order term as in the case of $K\to 2\pi$ decay mentioned above~\cite{CT}.}. This effect is missing in the calculations in S$\chi$EFT$_\pi$ available in the literature. It surely must be very important in nuclear correlations.

Now coming to the Ferm-liquid fixed point result (\ref{qL}), I would argue that $\Delta$-hole configurations should not figure in the quasiparticle structure involved in the calculation. This is because in this formulation the limit $1/\bar{N}\to 0$ is appropriate and the $m_\Delta -m_N\sim 300$ MeV is not relevant in RG sense.

Let me next address the question as to how to calculate the Gamow-Teller matrix element in neutrinoless double $\beta$-decay process where the kinematics involved is quite different from the Gamow-Teller transition considered up to here where zero momentum is involved. Here the situation is even more unfavorable for calculating the 2BC reliably. The reason is that the momentum transfer can be of order $\sim 100$ MeV, therefore it would make even less sense to stop at N$^3$LO. It could also be that the standard power expansion will break down, calling for a totally different power expansion.

Finally the ``mysterious" continuity of  $g_A^{\rm eff}\approx 1$ from low density to  high density. There is an analogous happening in compact stars. As suggested in \cite{MR-review},  there seems to be  a precocious emergence of pseudo-conformal symmetry in compact stars at a density $n\gsim 3n_0$ far below the asymptotic density where it is expected. On the other hand, at low density, the ``unitarity limit" associated with infinite scattering length is considered to be present in light nuclei~\cite{vankolck}  and also in low-density edge of compact stars~\cite{lattimer}. Now the point is that the unitarity limit leads to conformal invariance~\cite{wise}. Thus there is a hint of conformal symmetry permeating  from low density to high density. In between, in finite nuclei,  such a symmetry is invisible. It  may then be ``hidden" or ``lost." It would be amusing if the two continuities, both implicating scale symmetry, were related. This is a fascinating observation to be further investigated.

\end{document}